\documentclass[lettersize,journal]{IEEEtran}
\usepackage{amsmath,amsfonts}
\usepackage{algorithmic}
\usepackage{algorithm}
\usepackage{array}
\usepackage{multirow}
\usepackage{booktabs} 
\usepackage[caption=false,font=normalsize,labelfont=sf,textfont=sf]{subfig}
\usepackage{textcomp}
\usepackage{stfloats}
\usepackage{url}
\usepackage{verbatim}
\usepackage{graphicx}
\usepackage{cite}
\usepackage{bm}
\usepackage{amssymb}
\usepackage{diagbox}
\usepackage{float}
\begin{document}

\title{mmWave RIS Phase Shift Feedback Based on Knowledge Base Autoencoder Framework}

\author{Hao Feng, Yuting Xu, Yuping Zhao
%\thanks{This paper was produced by the IEEE Publication Technology Group. They are in Piscataway, NJ.}% <-this % stops a space
\thanks{(Corresponding author: Yuping Zhao.)}
\thanks{Hao Feng and Yuting Xu are with the Peking University Shenzhen Graduate School, Peking University, Shenzhen 518066, China, with the Peng Cheng Laboratory, Shenzhen 518066, China, and also with the School of Electronics, Peking University, Beijing 100871, China (e-mail: hfeng@pku.edu.cn, yutingxu@stu.pku.edu.cn).}
\thanks{Yuping Zhao is with the School of Electronics, Peking University, Beijing 100871, China (e-mail: yuping.zhao@pku.edu.cn).}}



\maketitle
\begin{abstract}
In reconfigurable intelligent surface (RIS) assisted wireless communication systems, adjusting the phase shift of RIS unit cells is crucial for improving communication performance. Due to massive RIS unit cells, the number of phase shift parameters fed back from the base station (BS) to the RIS is enormous, which occupies a large number of frequency resources. In this paper, we propose a feedback scheme for millimeter-wave RIS phase shift applying a knowledge base autoencoder framework, in which the learnable knowledge base is shared at the BS and the RIS. The encoder at the BS compresses the RIS phase shift matrix to multiple feature vectors. Then the knowledge base vectors index is obtained by calculating the similarity between feature vectors and knowledge base vectors and transmitted to the RIS. With utilizing the index at the RIS, the corresponding knowledge base vectors are extracted and used as the decoder’s inputs to reconstruct the phase shift of the RIS. Simulation results show that the proposed scheme can significantly improve the accuracy of phase shift feedback and impressively reduce the amount of RIS phase shift feedback data. Moreover, the proposed scheme is easy to deploy in actual scenarios due to lower complexity and fewer parameters.
\end{abstract}


\section{Introduction}
\IEEEPARstart{W}{ith} the large-scale commercial use of fifth-generation (5G), the 5G platform supports a broader range of applications requiring high-speed, massive connections, and ultra-reliable low-latency communication, such as high-resolution streaming media, virtual reality, and augmented reality~\cite{b01}, resulting in a considerable increase in data rates~\cite{b02,b03,b04,b05}. In order to meet new mobile application scenarios, new wireless communication standards need to migrate to higher frequencies, such as the millimeter wave frequency band (mmWave)~\cite{b06}. Although mmWave provides more bandwidths and services high data rate applications, it unignorably has severe path loss.

As a promising technology in next-generation wireless communication systems, reconfigurable intelligent surface (RIS)~\cite{b07} is used to enhance the coverage of mmWave signals and improve spectrum and energy efficiency with low power consumption and low hardware cost~\cite{b08}. To realize the reconstruction of the wireless propagation environment, RIS uses many low-cost passive components to intelligently control the reflection phase of the incident electromagnetic wave~\cite{b09}. When the line-of-sight link between the user equipment (UE) and the base station (BS) is blocked~\cite{b10}, RIS can improve the link quality of the UE~\cite{b11} and communication system performance~\cite{b12} by rationally configuring the phase shift of the RIS unit cells.

The phase control of RIS unit cells is a pivotal technology for RIS-assisted wireless communication system. In order to achieve better performance, the RIS-assisted wireless communication system needs to find the optimal phase shift in the cascaded channel and feed it back to the RIS through the feedback channel. Due to a large number of RIS unit cells, the vast phase shift parameters that need to be fed back take up a lot of frequency bands and time resources. It is difficult to achieve in rate-limited situations. Therefore, how to feed back accurately and reduce the phase shift parameters is an urgent problem to be solved. This issue has been neglected in existing RIS research work. To our knowledge, only one piece of literature studies this issue. A scheme based on a convolutional autoencoder is proposed in~\cite{b13}, which compresses the discrete phase shift by the encoder at the UE and reconstructs the discrete phase shift by the decoder at the RIS. However, this work feeds back discrete phase shifts and ignores the case of continuous phase shifts. Furthermore, sequentially encoding and decoding unit cell phase shift of RIS is inefficient.

In this paper, a novel mmWave phase shift feedback scheme is proposed for RIS based on a knowledge base autoencoder (KBAE) framework. The scheme considers the uplink communication system scenario. The BS needs to feed back the continuous phase shift of the RIS with optimal system performance to the RIS through a downlink feedback channel. Then the RIS uses the feedback phase shift to perform passive beamforming, and the signal sent by the UE will be focused on the BS after passing through the RIS. In the phase shift feedback, an encoder and a decoder are deployed at the BS and the RIS, respectively, with a learnable knowledge base shared at both ends. The encoder compresses the RIS phase shift matrix and outputs multiple feature vectors. Then the index of the knowledge base vectors which are most similar to the feature vectors is obtained and transmitted to the RIS. The RIS takes out the vectors corresponding to the index from the knowledge base and reconstructs the phase shift of the RIS through the decoder. As far as we know, no one has paid attention to the problem of RIS continuous phase shift matrix feedback. In addition, we first propose a KBAE framework to solve the mmWave RIS phase shift feedback problem.
% figure 01
\begin{figure}[!t]
\centering
\includegraphics[width=3.2in]{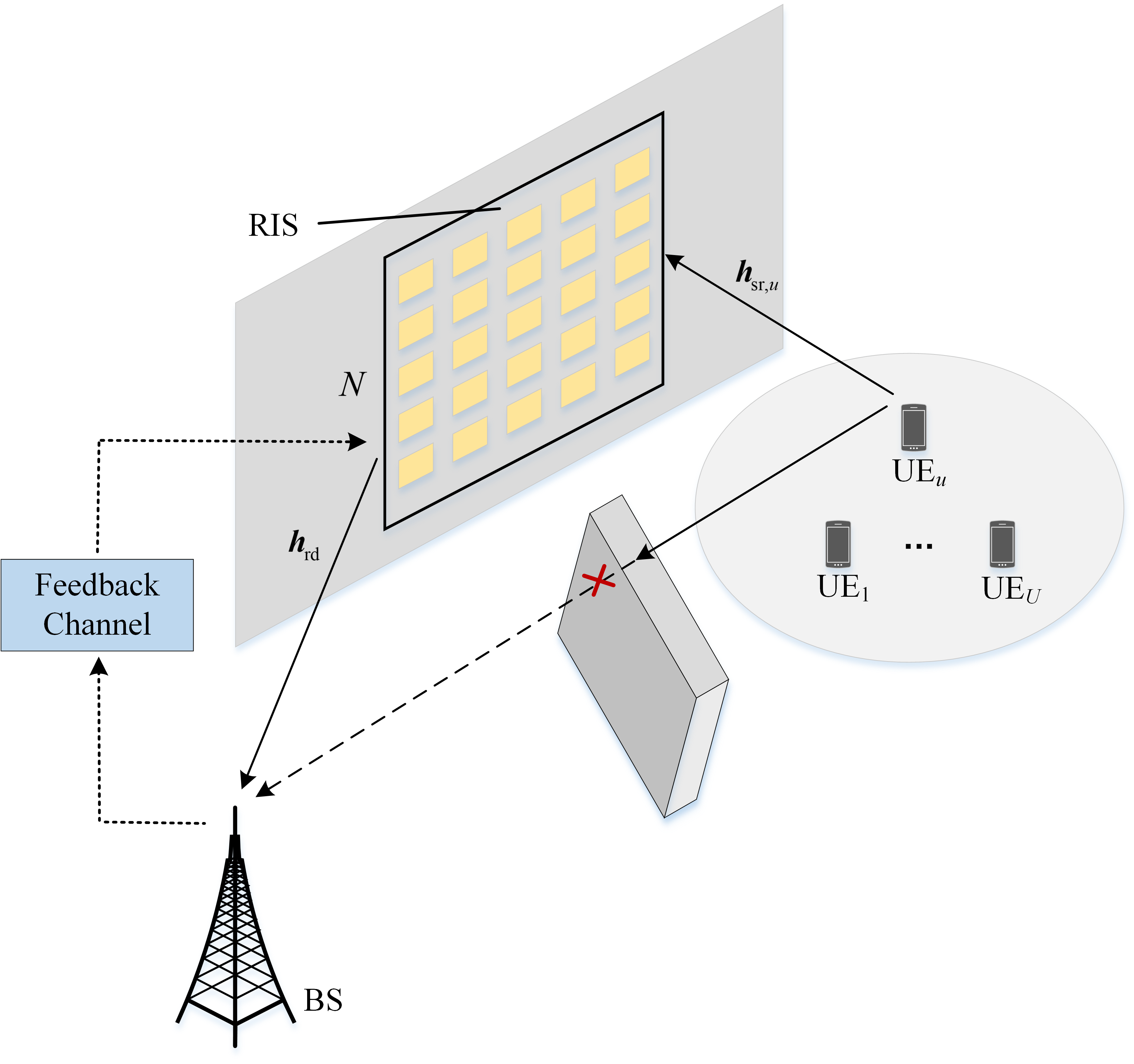}
\caption{Schematic diagram of the system model.}
\label{fig01}
\end{figure}

To summarize, the main contributions and innovations of this paper are as follows:
\begin{itemize}
\item A KBAE framework and a mmWave RIS phase shift feedback scheme based on the framework are proposed, which can significantly reduce the number of feedback phase shift parameters.
\item For different compression ratios, we propose two autoencoder network structures in the KBAE framework. And the global attention mechanism is introduced for a higher compression ratio to compress the network structure without losing reconstruction accuracy.
\item A method for establishing a learnable knowledge base is proposed. The knowledge base can learn the distribution of the encoder output feature vectors better to characterize feature information of the phase shift matrix.
\item The proposed RIS phase shift feedback scheme is simulated. The simulation results show that when the size of the knowledge base is reasonable, the RIS phase shift feedback accuracy of the scheme is better than the benchmark methods under different compression ratios.
\item The proposed scheme can feed back the RIS continuous phase shift matrix. At the same time, the network's time and space complexities are significantly reduced compared with the benchmark methods, thus improving the efficiency of RIS phase shift feedback.
\end{itemize}

The rest of this paper is organized as follows. Section II introduces the system model. Section III presents the proposed mmWave RIS phase shift feedback scheme based on the KBAE framework. Section IV introduces two autoencoder network structures in the designed KBAE framework. Section V is the simulation results, and Section VI summarizes the whole paper.

\section{System Model}
An uplink multi-user communication system is shown in Fig.~\ref{fig01}. There are $U$ UEs, and the BS and each UE configure an antenna. The RIS has $N$ passive reflection unit cells, and the number of unit cells in the horizontal and vertical directions is $M = \sqrt N $. Each unit cell can induce an independent phase shift to the reflected signal. Due to the severe path loss of the mmWave, the direct path signal from UE to BS is feeble, thus the direct path is not considered. Assuming that the system works in time-division duplex mode, the downlink channel can be estimated from the uplink channel according to channel reciprocity.

\subsection{Channel model}
Assume that all channels adopt a quasi-static block fading model, and we focus on communication in a specific fading block. As shown in Fig.~\ref{fig01}, ${\pmb{h}_{{\text{sr,}}u}} \in {\mathbb{C}^{N \times 1}}$ represents the channel vector from the $u$-th UE to the RIS and ${\pmb{h}_{{\text{rd}}}} \in {\mathbb{C}^{N \times 1}}$ represents the channel vector from the RIS to the BS. $\pmb{\Omega}  = {\text{diag}}\left\{ {{\phi _1},{\phi _2}, \ldots ,{\phi _N}} \right\} \in {\mathbb{C}^{N \times N}}$ represents the reflection coefficient matrix of RIS, where ${\phi _i}{ = }\alpha {e^{j{\theta_i}}}$, $i \in \left\{ {1, \ldots ,N} \right\} \buildrel \Delta \over = {\cal N}$, $\alpha  \in \left( {0,1} \right]$ is the reflection coefficient amplitude, and ${\theta _i} \in \left[ {0,2{\rm{\pi }}} \right)$ is the reflection phase shift. In this paper, it is assumed that the magnitude of the RIS reflection coefficient $\alpha =$ 1.

\subsection{Signal model}
In the uplink channel, the equivalent channel from the $u$-th UE to the BS can be expressed as
% equation-01
\begin{align}
{\pmb{h}_u} &= \pmb{h}_{{\text{rd}}}^T\pmb{\Omega} {\pmb{h}_{{\text{sr}},u}}  \\
 &= {\pmb{H}_u}\pmb{\phi}, \notag
\end{align}
where $\pmb{\phi}  \buildrel \Delta \over = {\left[ {{e^{j{\theta _1}}}, \ldots ,{e^{j{\theta _N}}}} \right]^T}$ represents the phase shift vector of the RIS and ${\pmb{H}_u} \triangleq \pmb{h}_{{\text{rd}}}^T{\text{diag}}\left( {{\pmb{h}_{{\text{sr}},u}}} \right) \in {{\mathbb C}^{1 \times N}}$ represents the cascaded channel from the $u$-th UE to the BS. Since we only consider the phase shift feedback, we assume that ${\pmb{h}_{{\text{rd}}}}$ and ${\pmb{h}_{{\text{sr}},u}}$ are known. According to the channel capacity maximization criterion, the optimal phase shift at the BS corresponding to the $u$-th UE can be expressed as~\cite{b14}
% equation-02
\begin{equation}
\label{eq-2}
{\left[ {\theta _i^ * } \right]_u} = \arg \left( {{{\left[ {{\pmb{h}_{{\text{sr}},u}}} \right]}_i}{{\left[ {{\pmb{h}_{{\text{rd}}}}} \right]}_i}} \right),i \in {\cal N}.
\end{equation}

The BS needs to feed back the optimal RIS phase shift of the $u$-th UE to the RIS. When the $u$-th UE sends a signal, the RIS adjusts the reflection phase of its unit cells in accordance with the feedback phase shift to focus the signal on the BS, thereby improving the communication performance of the $u$-th UE. A difficulty here is that the number of unit cells of RIS is usually large, generally hundreds or even thousands. Moreover, the phase value of each unit cell needs a high quantization precision. As a result, the number of bits is enormous after analog to digital conversion. A tremendous amount of data needs to be fed back, which occupies many spectrum resources and generates a considerable delay. Therefore, reducing the amount of data in the feedback phase shift is an urgent problem to be solved.
% figure 02
\begin{figure*}[!t]
\centering
\includegraphics[width=5.6in]{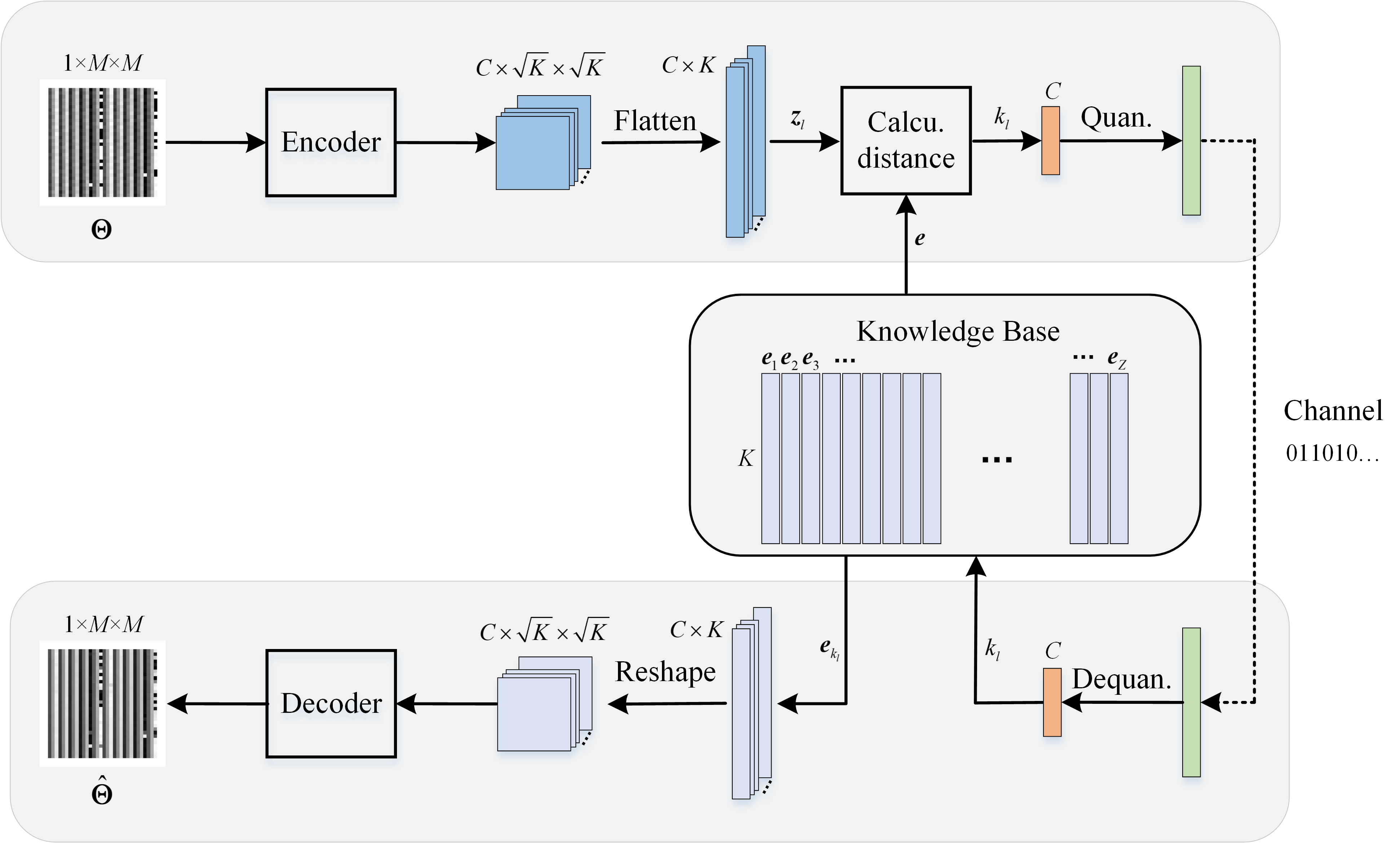}
\caption{Flow chart of RIS phase shift feedback based on KBAE framework.}
\label{fig02}
\end{figure*}

\section{RIS Phase Shift Feedback Based on KBAE Framework}
This paper proposes a mmWave RIS phase shift feedback scheme based on the KBAE framework to reduce the feedback overhead. The KBAE framework includes two parts: the knowledge base and the autoencoder. This section first introduces the proposed KBAE framework and RIS phase shift feedback process, then introduces the construction of the knowledge base and quantization, and finally introduces the loss function for the KBAE framework. The designed autoencoder structures that can be used in the KBAE framework will be introduced in detail in Section IV.

\subsection{RIS Phase Shift Feedback Process Based on KBAE Framework}
Fig.~\ref{fig02} is a flow chart of mmWave RIS phase shift feedback based on KBAE. An encoder and a decoder are deployed respectively at the BS and RIS. Firstly, the encoder compresses the RIS phase shift and generates multiple feature vectors. Then the index of the knowledge base vectors most similar to the feature vectors is obtained by calculating the similarity between the feature vectors and the pre-learned knowledge base vectors. Next, the index is converted into a bit stream and transmitted to the RIS. When receiving the index correctly, the RIS takes out the vector from the knowledge base with the index. Finally, the phase shift of the RIS can be reconstructed by the decoder. With a learnable and shared knowledge base, only the indexes of knowledge base vectors need to be transmitted instead of a large number of phase shift parameters, dramatically reducing the occupation of frequency band resources.

The RIS phase shift feedback process based on the KBAE framework can be divided into three stages.

\textbf{1) Encoder compression}: We first reshape the $u$-th UE optimal phase shift vector (2) to a matrix $\Theta $. The encoder extracts the high-dimensional features of the phase shift matrix $\Theta $ and outputs $C$ feature matrices. Then flatten the feature matrices into feature vectors ${\pmb{z}_l} \in {{\mathbb C}^{K \times 1}}$, $l \in \left\{ {1,2, \ldots ,C} \right\} \buildrel \Delta \over = {\mathcal C}$,
% equation-03
\begin{equation}
\label{eq-3}
{\pmb{z}_l} = {\rm{Flatten}}\left( {{f_{{\rm{encode}}}}\left( \Theta  \right)} \right),
\end{equation}
where ${f_{{\text{encode}}}}\left(  \cdot  \right)$ and Flatten$(\cdot)$ represent the compression model and flattening operation, respectively, and the length of the feature vector is the same as the length of the knowledge base $K$. To be noted, $C$ is related to the amount of system feedback data, which has a practical physical meaning and is equal to the number of encoder output and the decoder input convolution channels.

\textbf{2) Using the knowledge base to compress further}: We measure the similarity between the feature vectors ${\pmb{z}_l}$ and the knowledge base vectors ${\pmb{e}_i} \in {{\mathbb C}^{K \times 1}},i \in \left\{ {1,2, \ldots ,Z} \right\} \triangleq { \cal Z}$, where $Z$ is the size of the knowledge base, and obtain the index ${k_l}$ of the knowledge base vectors ${\pmb{e}_{{k_l}}}$ which are most similar to ${\pmb{z}_l}$. Then, ${k_l}$ is quantized to form a bit stream, which needs to be transmitted to the RIS through a feedback channel,
% equation-04
\begin{equation}
\label{eq-4}
k_l^{{\rm{bit}}} = Q\left( \mathop {\arg \min }\limits_{i \in {\cal Z}} {Dis\left( {{\pmb{z}_l},\pmb{e}_i} \right)} \right),
\end{equation}
where $Dis(\cdot)$ and $Q(\cdot)$ denote similarity measure and quantization, respectively, and $k_l^{{\rm{bit}}}$ is the bit form of ${k_l}$. The construction of the knowledge base and quantification will be detailed below.

\textbf{3) Decoder reconstruction}: Assuming that $k_l^{{\rm{bit}}}$ can be received correctly at the RIS, the RIS dequantizes the received signal to acquire index ${k_l}$,
% equation-05
\begin{equation}
\label{eq-5}
{k_l} = Deq\left( {k_l^{{\rm{bit}}}} \right),
\end{equation}
where $Deq(\cdot)$ represents the dequantization. Then the corresponding vectors ${\pmb{e}_{{k_l}}}$ are taken out from the knowledge base according to ${k_l}$. Finally, ${\pmb{e}_{{k_l}}}$ is converted into a matrix and input to the decoder, and the decoder reconstructs the phase shift matrix
% equation-06
\begin{equation}
\label{eq-6}
\hat \Theta  = {f_{{\rm{decode}}}}\left( {{\rm{Reshape}}\left( {{\pmb{e}_{{k_l}}}} \right)} \right),
\end{equation}
where ${f_{{\rm{decode}}}}\left(  \cdot  \right)$ and Reshape$(\cdot)$ denote decompression model and reshaping operation, respectively.

Owing to the excellent performance of neural networks in compression and restoration, we use neural networks as encoders and decoders. Aided by the knowledge base, the proposed KBAE framework transforms the original element-by-element transmission of compressed features into the transmission of the knowledge base vectors index. As long as the knowledge base represents the feature distribution better, the KBAE framework reduces the amount of transmitted data and achieves better reconstruction accuracy. In addition, the KBAE framework can freely use different autoencoder structures for different tasks and scenarios. We propose two autoencoder structures for different tasks, which will be introduced in detail in Section IV.

\subsection{Construction of Knowledge Base and Quantification}
The knowledge base is the knowledge content shared between the BS and the RIS, and its dimension is $\pmb{e} \in {{\mathbb C}^{K \times Z}}$, where $K$ is the length of each knowledge base vector ${\pmb{e}_i}$, and $Z$ is the size of the knowledge base, that is, there are $Z$ embedding vectors ${\pmb{e}_i}$ stored in the knowledge base. The knowledge base is set to a uniform distribution of (0, 1$/K$) in the network initialization phase. The initialization is to control the model loss from being too large during the knowledge base learning phase, thereby accelerating model training and convergence.

The encoder outputs $C$ $K$-dimensional feature vectors ${\pmb{z}_l}$. The index ${k_l}$ of the knowledge base vectors is obtained by calculating the similarity between knowledge base vectors and ${\pmb{z}_l}$, as shown in the following formula
% equation-07
\begin{equation}
\label{eq-7}
{k_l} = \mathop {\arg \min }\limits_{i \in {\cal Z}} {\left\| {{\pmb{z}_l} - {\pmb{e}_i}} \right\|_2^2},l \in {\cal C}.
\end{equation}

The index ${k_l}$ needs to be quantized to form a bit stream to be transmitted. Because the quantized index needs to represent the entire knowledge base, the number of quantization bits $q$ is related to the size of the knowledge base $Z$, which can be expressed as
% equation-08
\begin{equation}
\label{eq-8}
q = {\log _2}\left( Z \right).
\end{equation}
Moreover, the system only needs to transmit $C$ knowledge base vector indexes, which correspond to $C$ feature vectors output by the encoder. The total number of bits to be transmitted is
% equation-09
\begin{equation}
\label{eq-9}
B = q \times C = {\log _2}\left( Z \right) \times C.
\end{equation}
Then, it was originally necessary to transmit $N$ unit cells phase shifts, while only $C$ indexes need to be transmitted now so that the compression ratio can be expressed as
% equation-10
\begin{equation}
\label{eq-10}
\gamma  = \frac{N}{C}.
\end{equation}

Assume that the bit stream sent by the BS is correctly received at the RIS, and $C$ indexes ${k_l}$ are generated after dequantization. Then with the index ${k_l}$ the corresponding vectors from the knowledge base are extracted as the input of the decoder
% equation-11
\begin{equation}
\label{eq-11}
{\pmb{z}'_l} = {\pmb{e}_{{k_l}}},l \in {\cal C}.
\end{equation}
Finally, the phase shift is reconstructed by the decoder at the RIS.

The encoder transforms the input phase shift matrix into feature vectors forming a phase shift feature space, and the knowledge base uses a limited number of vectors to approximate the phase shift feature distribution. When the size of the knowledge base $Z$ is increased, the knowledge base vectors approximate better to the phase shift feature distribution, consequently improving the reconstruction accuracy. In addition, expanding the dimension of the phase shift feature vector legitimately retains more phase shift feature information. With correspondingly increasing the dimension of the knowledge base, the distribution of the knowledge base vectors approaches the phase shift feature distribution more flexibly, thereby further improving the reconstruction accuracy.

%%---------- Loss Function for KBAE framework--------- %%
\subsection{Loss Function for KBAE framework}
We take inspiration from~\cite{b15} and use (12) to represent the overall loss function which consists of two parts. The first term is the mean squared error (MSE) loss between the encoder's input optimal phase shift $\Theta $ and the decoder's output reconstructed phase shift $\hat \Theta $, representing the reconstruction loss. The second term is the MSE loss between the knowledge base vectors and the encoder output feature vectors, which is used to represent the learning of the knowledge base, denoted as $kb$ loss. The overall loss function can be expressed as
% equation-12
\begin{equation}
\label{eq-12}
Loss = \left\| {\Theta  - \hat \Theta } \right\|_2^2 + k{b_{{\rm{loss}}}}.
\end{equation}

The $kb$ loss can be divided into two items which can be expressed as
% equation-13
\begin{equation}
\label{eq-13}
k{b_{{\rm{loss}}}} = {\mathop {\sum }\limits_{l \in {\cal C}}} \left ( \left\| {sg[{\pmb{z}_l}] - \pmb{e}_{{k_l}}} \right\|_2^2 + \beta \left\| {{\pmb{z}_l} - sg[\pmb{e}_{{k_l}}]} \right\|_2^2 \right ),
\end{equation}
where $sg[\cdot]$ represents the stop gradient operator with equivalent and zero gradients during forward calculation so that the operand can be effectively constrained to be a non-updating constant. $\beta $ acts as a penalty factor to control the gradient update rates of the knowledge base and encoder in $kb$ loss. The former item in the $kb$ loss is set to train the knowledge base, which makes the vector distribution in the knowledge base closer to the encoder output vector distribution. The latter term is used to fix the knowledge base so that the trainable encoder output vector distribution approximates the knowledge base vector distribution. The $kb$ loss makes it promising that the distribution of the knowledge base vectors is closer to the encoder output vector distribution.
% figure 03
\begin{figure*}[!t]
\centering
\includegraphics[width=5.5in]{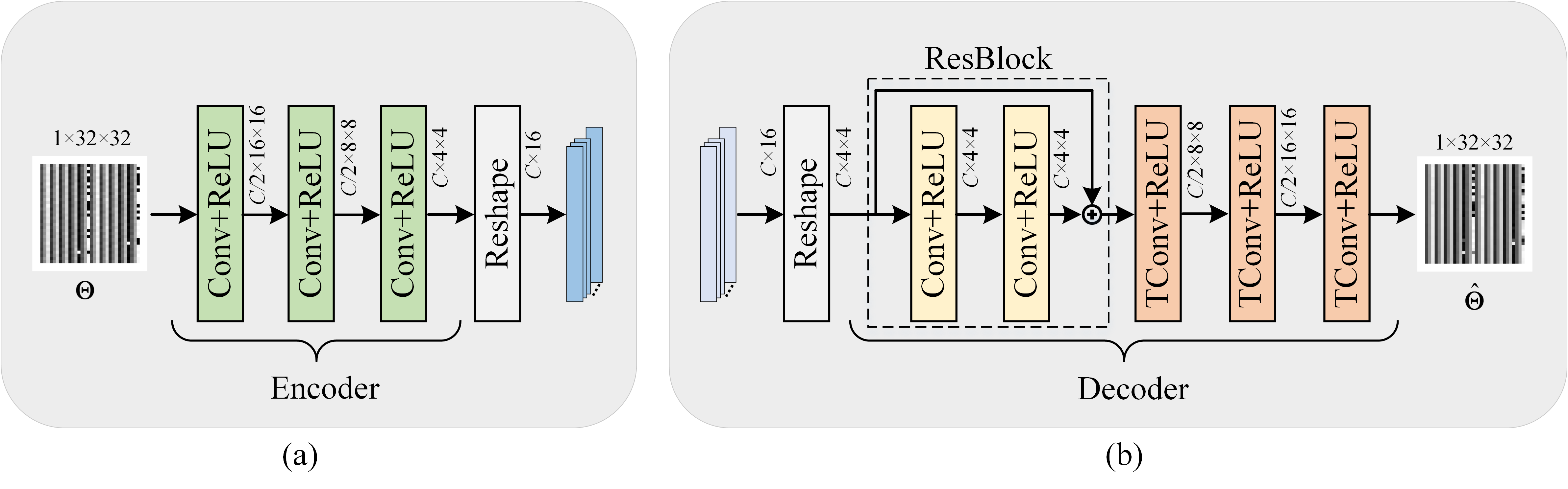}
\caption{PSFNet encoder and decoder structures.}
\label{fig03}
\end{figure*}

\section{Autoencoder Network For the KBAE Framework}
The autoencoder structure in the KBAE framework can use different models applied to multiple tasks and scenarios to achieve better performance. In this paper, we propose two different autoencoder structures, phase shift feedback network (PSFNet) and PSFNet for a higher compression ratio (PSFNet-H).
\subsection{PSFNet Autoencoder Structure}
Neural networks are widely used in image compression and restoration tasks and have achieved impressive results. Therefore, the encoder and the decoder in the PSFNet autoencoder use a neural network architecture. Taking $M=$ 32 as an example to introduce the structure of the PSFNet in detail:

\textbf{Encoder}: The encoder's input is the normalized optimal phase shift matrix $\Theta $, and the dimension is 1$\times$32$\times$32. Moreover, it passes through 3 combined layers of convolution and ReLU sequentially, and the kernel size of each convolution layer is 4$\times$4, the stride is 2, and the padding is 1. Each designed convolution layer compresses the feature matrix to its 1/4 so that the full convolution module can extract features and compress the matrix simultaneously. Using a fully convolutional module instead of a conventional fully connected module for compression can reduce model complexity and promote model generalization and robustness. The number of convolution channels is $C/$2, $C/$2, and $C$, respectively. After a flattening operation, the encoder's output is the feature vectors of $C \times $16. These feature vectors will be subject to a similarity measure with the vectors in the knowledge base.

\textbf{Decoder}: The input of the decoder is $K=$ 16-dimensional vectors taken from the knowledge base, then converted into matrices through reshaping operation. First, it goes through a residual module containing two layers of convolution and ReLU. The convolution kernel size is 3$\times$3, the stride is 1, and the padding is 1. Adding the residual module to the decoder makes information's forward and backward propagation smoother. Since the residual module contains an identical natural mapping, it solves the problems of network degradation and gradient dispersion to a certain extent. After the residual module, the rest of the decoder is designed to be symmetrical to the encoder. There are three upsampling layers composed of transposed convolution (TConv) and ReLU, in which the kernel size of the TConv is 4$\times$4, the stride is 2, the padding is 1, and the number of convolution channels is $C/$2, $C/$2, and 1. The output of the decoder is denoted as $\hat \Theta $.

\subsection{PSFNet-H Autoencoder Structure}

Under a specific feedback accuracy requirement in phase shift feedback, the fewer bits fed back, the better. Besides, the model cannot be too large, which shortens the network running time and improve generalization performance. Therefore, in this section, we propose a lightweight network PSFNet-H for a higher compression ratio, which performs higher compression on phase shift while the network is lighter and achieves competitive feedback accuracy requirements. PSFNet-H is mainly designed for high compression ratio, light weight, and high feedback accuracy requirements. Compared with PSFNet, we modified the encoder and decoder's network structure and the length $K$ of the knowledge base vector.

In the encoder and decoder design, PSFNet-H increases the dimension of the encoder output feature vectors compared with PSFNet, so the encoder output carries more phase shift information. At the same time, because the high compression ratio means reducing the amount of transmitted data, the transmitted data must carry richer information as much as possible to make up for information loss, that is, the transmitted index of the knowledge base vector must contain as much information as possible. Therefore, we apply a global attention residual module (GARB). The global attention is used to strengthen the multi-channel information fusion of the feature matrix and the residual can strengthen the information fusion of forward and backward propagation. Applying the GARB enables our designed network to handle high compression ratio tasks sophisticatedly. In detail, we will take $M=$ 32 as an example to introduce the structure of PSFNet-H.

\textbf{Encoder}: The encoder's input is also the optimal phase shift matrix $\Theta $, with a dimension of 1$\times$32$\times$32, which first passes through a combination layer of convolution and ReLU. Passing through the GARB, the output feature matrix’s dimension is 8$\times$16$\times$16. Next, continue passing through the combined layer of convolution and ReLU. The kernel of the first two combined layers is 4$\times$4, the stride is 2, the padding is 1, and the number of convolution channels is 8. After another GARB, the output dimension is 8$\times$8$\times$8. Then continue going through a combined layer of convolution and ReLU. The kernel of the last layer is 3$\times$3, the stride is 1, the padding is 1, and the number of convolution channels is $C$. Finally, after a flattening operation, the output vector dimension is $C$$\times$64.

\textbf{Decoder}: The input of the decoder is $K=$ 64-dimensional vectors extracted from the knowledge base, converted into matrices through reshaping operations. After a GARB, the output dimension is $C\times$8$\times$8. Then after an upsampling layer composed of TConv and ReLU, the output dimension is 8$\times$16$\times$16. Next, going through a GARB, the output dimension has no change. Finally, the feature matrix goes through the upsampling layer mentioned before. The kernel size of the TConv in the upsampling layer is 4$\times$4, the stride is 2, and the padding is 1. The decoder output is denoted as $\hat \Theta $.

\textbf{GARB module}: Fig.~\ref{fig05} shows the composition of the GARB module. GARB is a two-branch network and includes two parts: global attention mechanism and residual learning. The global attention mechanism collects complete information from each feature matrix channel to acquire more abundant features. In addition, residual is introduced as a connection from GARB’s input to output, in which the identical mapping significantly reduces the calculation complexity and improves the network's performance.
% figure 04
\begin{figure*}[!t]
\centering
\includegraphics[width=5.5in]{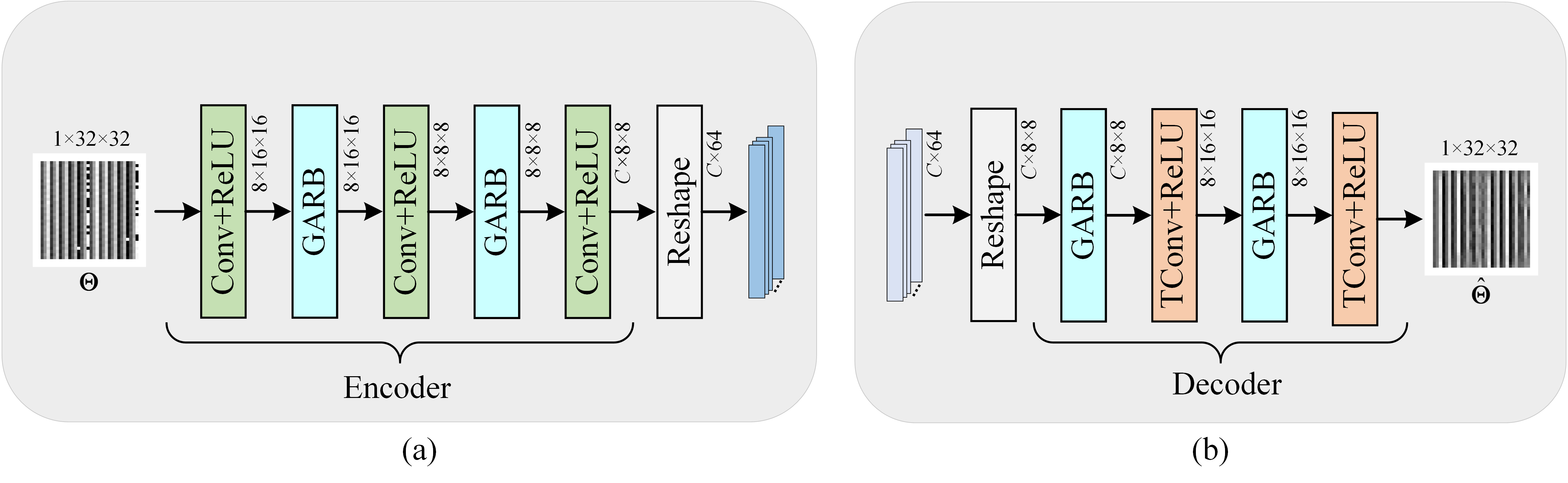}
\caption{PSFNet-H encoder and decoder structures.}
\label{fig04}
\end{figure*}

\textbf{a) Global attention mechanism}: The module input ${b_{i - 1}}$ passes through the 3$\times$3 convolutional layer, ReLU, and another 3$\times$3 convolutional layer. The middle output is ${R_i}$, whose dimension is $H \times W \times C$, where $H$ is height, $W$ is width, and $C$ is the number of convolution channels. After the global pooling layer, the output dimension is 1$\times$1$\times C$. Then the 1$\times$1 convolutional layer replaces the~\cite{b16} fully connected layer to reduce the number of model parameters, and the output is 1$\times$1$\times C / {{k_0}}$, where ${k_0}$ is the global scaling factor. After ReLU and another 1$\times$1 convolutional layer, the output dimension is restored to 1$\times$1$\times C$, and the output of the sigmoid is marked as ${S_i}$. ${R_i}^\prime $ denotes the multiplication output of ${S_i}$ and ${R_i}$, and its dimension is $H\times W \times C$. The global attention mechanism makes the output features integrate information from multiple channels, that is, each output channel collects more feature information from other channels, which meets the needs of a high compression ratio, and has the advantages of lightweight, high robustness, and better model performance.

\textbf{b) Residual learning}: Residual learning is introduced to ensure training stability and avoid problems such as gradient dispersion. The output ${R_i}^\prime $ after feature fusion is finally added to the input ${b_{i - 1}}$ to obtain the GARB output ${b_i}$
% equation-14
\begin{equation}
\label{eq-14}
{b_i} = {R_i}^\prime {\rm{ +  }}{b_{i - 1}}.
\end{equation}

Due to the increase of the compression ratio, the system can transmit less information, that is, the number of indexes transmitted becomes less. To convey more information about the phase shift, we need to increase the length of the knowledge base so that the distribution of the knowledge base vectors is more similar to the encoder output distribution. Therefore, the knowledge base vectors represent the encoder output vectors better, which improves the phase shift reconstruction accuracy. Here we increase the length of the knowledge base $K$ in PSFNet-H, equivalent to increasing the information carried by each knowledge base vector, which helps to promote the reconstruction accuracy of phase shift under a higher compression ratio.
% figure 05
\begin{figure}[!t]
\centering
\includegraphics[width=2.2in]{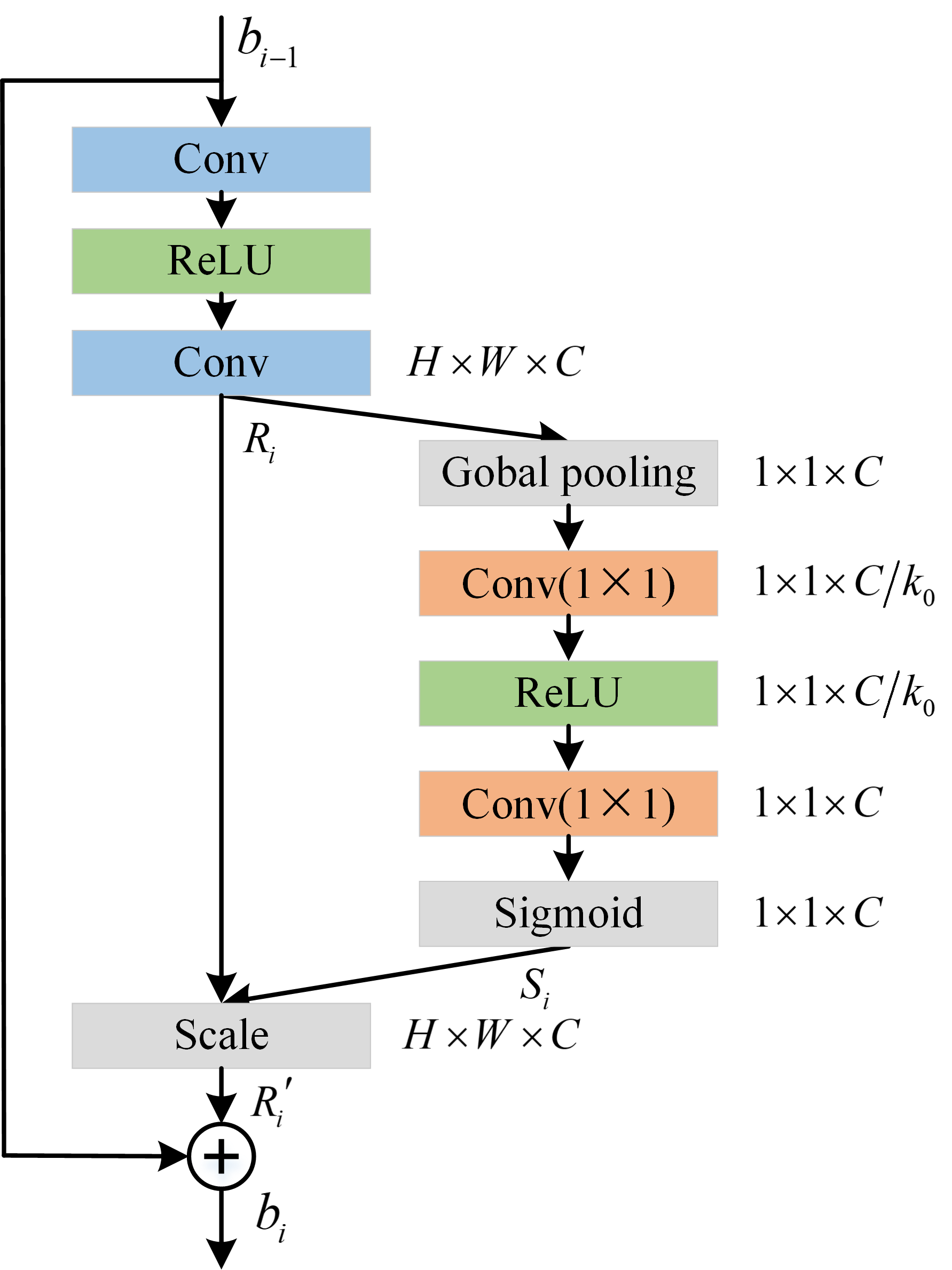}
\caption{ The $i$-th GARB module.}
\label{fig05}
\end{figure}
\section{Simulation Results}
In this section, to verify the performance of our proposed mmWave RIS phase shift feedback scheme based on the KBAE framework, we use the normalized mean squared error (NMSE) between the reconstructed phase shift $\hat \Theta $ and the optimal phase shift $\Theta $ sent by the BS as the performance index which can be expressed as
% equation-15
\begin{equation}
\label{eq-15}
{\text{NMSE}} = \frac{{{\mathbb E}\left[ {\left\| {\hat \Theta  - \Theta } \right\|_2^2} \right]}}{{{\mathbb E}\left[ {\left\| \Theta  \right\|_2^2} \right]}}.
\end{equation}

The $u$-th UE optimal phase shift $\Theta $ is calculated from channel vectors ${\pmb{h}_{{\text{rd}}}}$ and ${\pmb{h}_{{\text{sr}},u}}$. The channel vectors ${\pmb{h}_{{\text{rd}}}}$ and ${\pmb{h}_{{\text{sr}},u}}$ are generated by the widely used SimRIS channel simulator developed by KOC University ~\cite{b17}. The simulator considers the urban microcell mmWave narrowband channel model with operating frequencies at 28 GHz and 73 GHz (the scenario of our scheme is the urban microcell with operating frequency at 28 GHz). Besides, the simulator supports RIS far-field terminations, which is a reasonable assumption for smaller RIS sizes and mmWave with shorter wavelengths. 

There are three comparison benchmarks for the feedback problem of massive MIMO systems: 

1) CRNet in~\cite{b18}: Improving reconstruction performance by extracting multi-resolution features; 

2) CLNet in~\cite{b19}: the spatial attention mechanism is applied to the network to obtain the global receptive field; 

3) CsiNet in~\cite{b20}: first proposed a channel feedback framework based on deep learning autoencoder.

\subsection{PSFNet Simulation Results}
For PSFNet, the simulation parameters are $N=$ 1024, $M=$ 32, $K=$ 16, $\beta =$ 0.25, $k_0=$ 2, the learning rate is 0.002, the batch size is 200, and the number of training epochs is 200. We generated a total of 72,000 channel samples and got 72,000 optimal RIS phase shifts $\Theta $, where the number of the training set, test set, and validation set is 60,000, 6,000, and 6,000, respectively. All models are implemented under the framework of Pytorch1.11, and Nvidia A40 GPU is used for training. The Adam optimizer is chosen for training. The following formula can express the cosine annealing strategy selected as the learning rate adjustment strategy
% equation-16
\begin{equation}
\label{eq-16}
{\eta _t} = {\eta _{\min }} + \frac{1}{2}\left( {{\eta _{\max }} - {\eta _{\min }}} \right)\left( {1 + \cos \left( {\frac{{{T_{{\rm{cur}}}}}}{{{T_{\max }}}}} \right)} \right),
\end{equation}
where ${\eta _t}$ is the learning rate of the current epoch, ${\eta _{\min }}$ is the minimum learning rate equal to 0, ${\eta _{\max }}$ is the maximum learning rate equal to 0.002, ${T_{{\rm{cur}}}}$ is the current epoch, and ${T_{\max }}$ is the maximum learning rate adjustment epoch 20.

% figure 06
\begin{figure}[!t]
\centering
\includegraphics[width=3.2in]{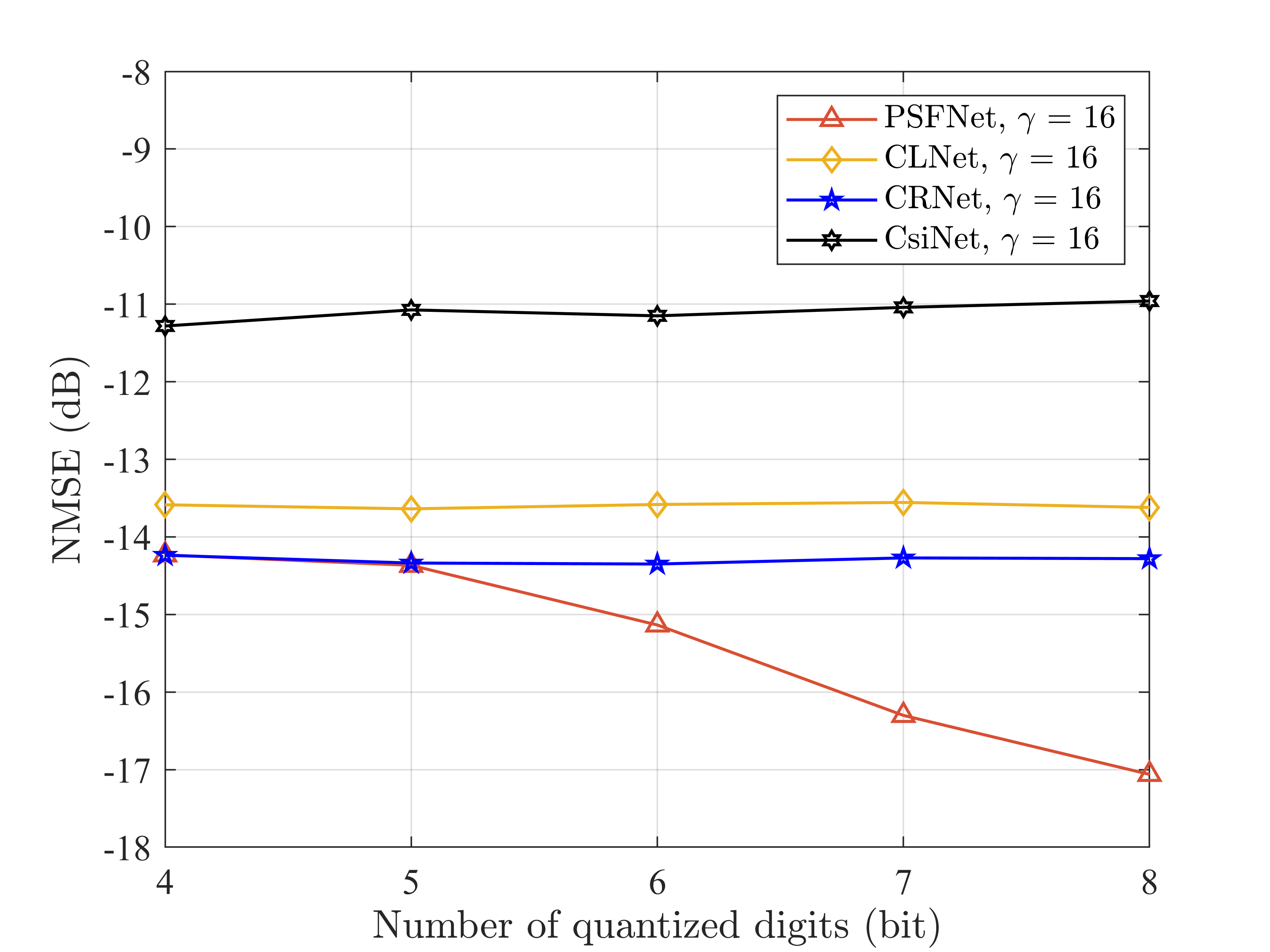}\\
(a)\\
\includegraphics[width=3.2in]{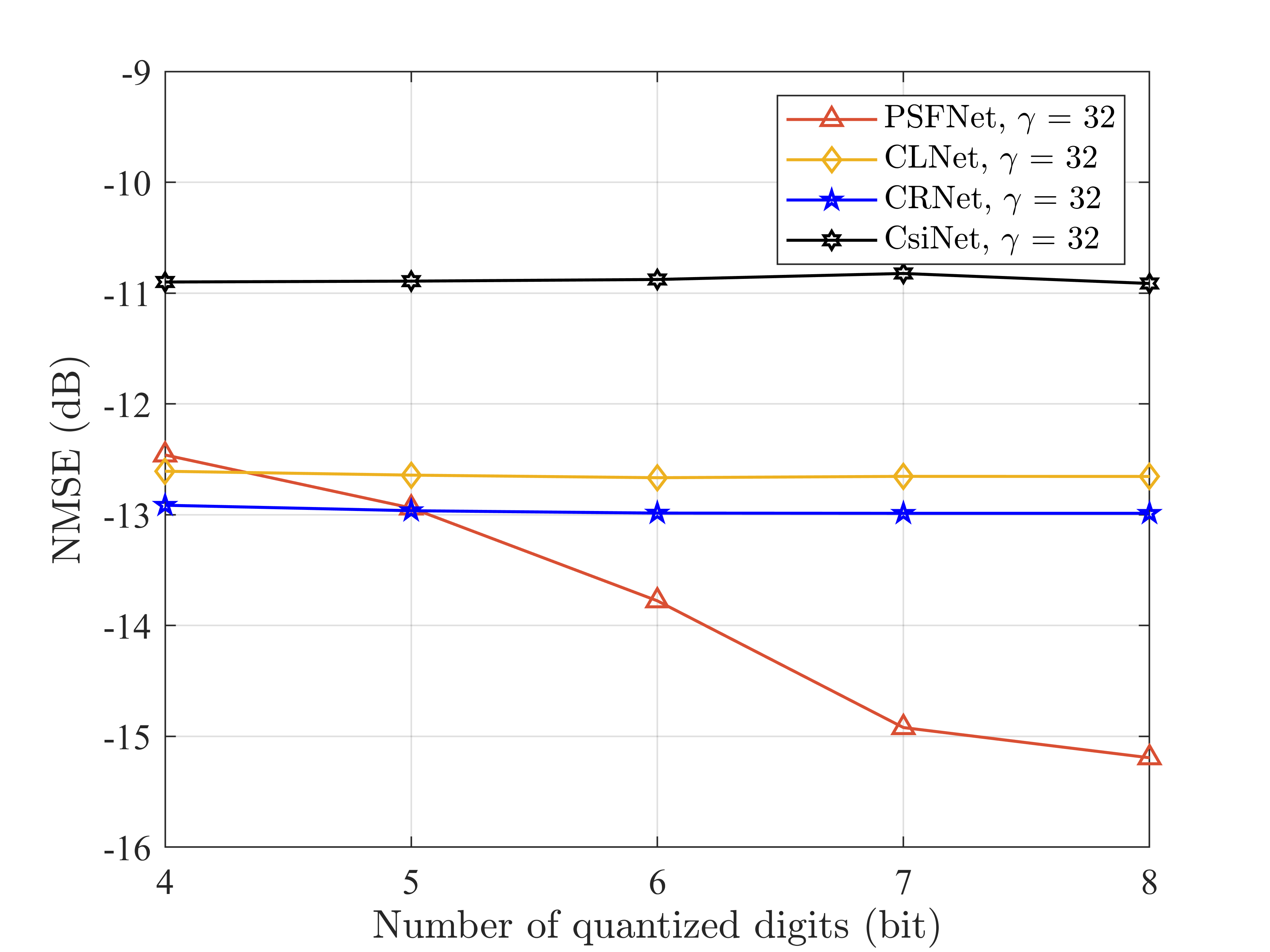}\\
(b)\\
\includegraphics[width=3.2in]{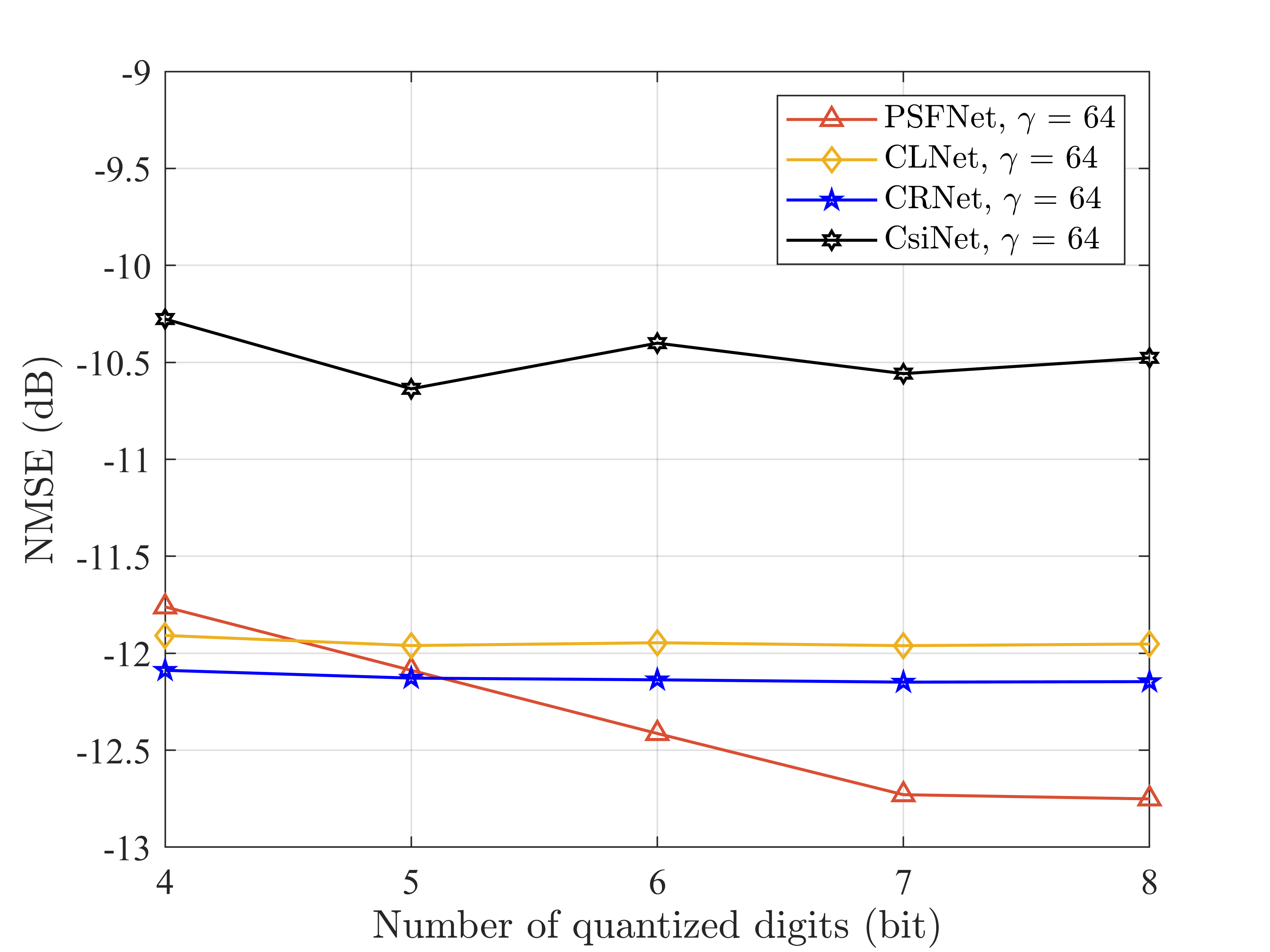}\\
(c)
\caption{Comparison of NMSE performance under different compression ratios.}
\label{fig06}
\end{figure}

Fig.~\ref{fig06} shows the NMSE performance comparison among PSFNet, CRNet, CLNet, and CsiNet under different compression ratios. In Fig.~\ref{fig06} (a), (b), and (c), compression ratios are successively set to $\gamma =$ 16, 32, 64. As shown in the figure, CsiNet has the largest NMSE and the worst performance in any case. The NMSE of PSFNet decreases with increasing the number of quantization bits, and only when the number of quantization bits is 4, the NMSE of PSFNet has a small gap compared with CLNet and CRNet. Once the number of quantization bits increases, the NMSE of PSFNet drops rapidly and is much smaller than the other three methods. It illustrates that as the number of quantization bits increases, the size of the knowledge base increases, and the phase shift information that the knowledge base can represent increases simultaneously, thus improving the accuracy of feedback phase shift reconstruction. According to Fig.~\ref{fig06}, it can be concluded that under different compression ratios, the performance of PSFNet improves significantly as the number of quantization bits increases. Because PSFNet is a model based on the KBAE framework, the number of quantized bits is equal to the size of the knowledge base. The larger the knowledge base is, the more phase shift information represents and the better the model performs. In addition, the main reason for the poor performance of PSFNet with quantization digits of 4 is that the knowledge base stores only 16 knowledge base vectors. The learned feature information is overly little, resulting in relatively poor model performance. The simulation results verify that the knowledge base needs a reasonable capacity to guarantee reconstruction accuracy and reduce transmission costs.
\begin{table*}[!t] 
\caption{Comparison of FLOPs and model size between PSFNet and other three benchmark networks.}
\centering
\begin{tabular}{cllllll}
\toprule
     Compression ratio   & \multicolumn{2}{c}{$\gamma = 16$} & \multicolumn{2}{c}{$\gamma = 32$} & \multicolumn{2}{c}{$\gamma = 64$}  \\
\midrule
     Model   & FLOPs & Model parameters & FLOPs & Model parameters & FLOPs & Model parameters \\
\midrule
 PSFNet (Ours)   & 9.703 M & 173.251 K & \bf{2.593 M (85.2\%)} 
               & \bf{43.619 K (64.1\%)} & \bf{733.184 K (24.4\%)}& \bf{11.059 K (31.4\%)}\\

 CRNet   & \bf{3.107 M} & 134.906 K & 3.041 M 
               & 69.338 K & 3.009 M & 36.554 K \\

CLNet   & 3.249 M & 135.032 K & 3.184 M 
               & 69.455 K & 3.151 M & 36.671 K \\

CsiNet   & 3.164 M & \bf{133.627 K} & 3.099 M
               & 68.059 K & 3.066 M & 35.275 K \\
\bottomrule
\end{tabular}
\label{tab1}
\end{table*}
% figure 07
\begin{figure*}[!t]
\centering
\includegraphics[width=6.2in]{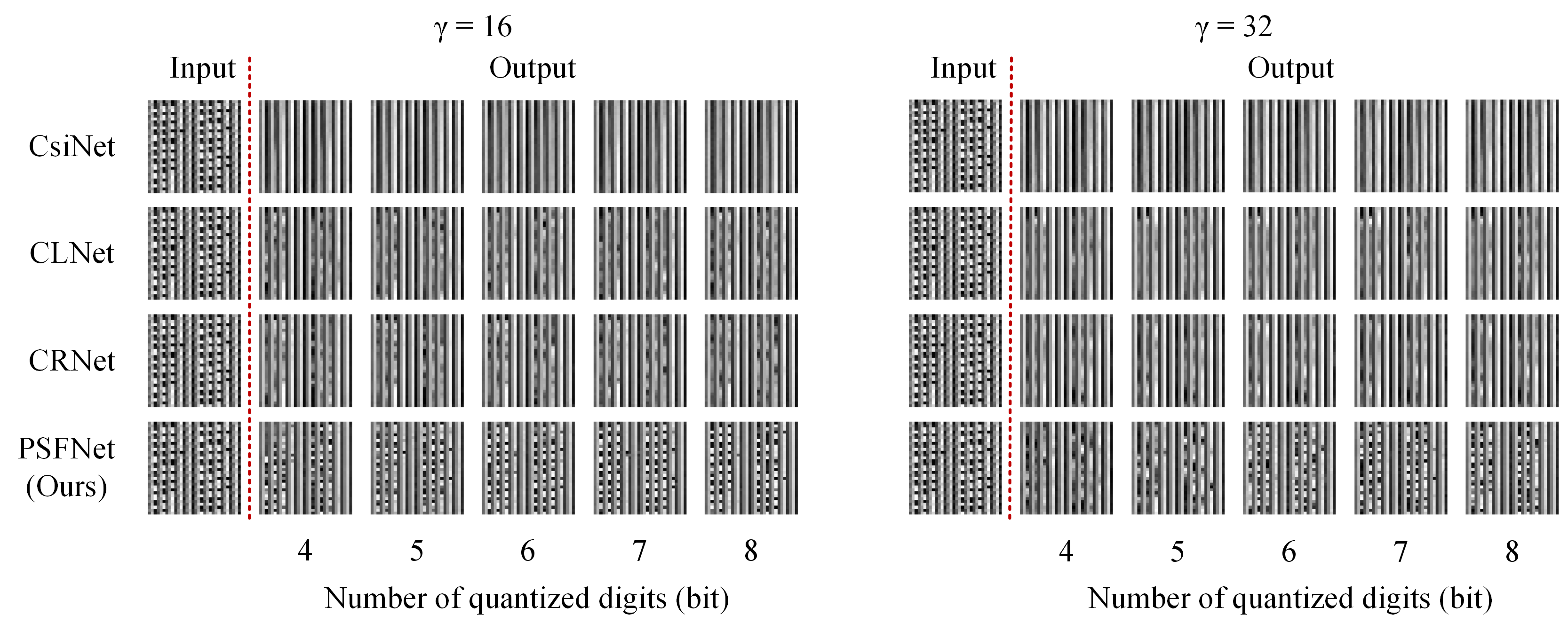}
\caption{Visualization of PSFNet and other three benchmark methods reconstructing phase shift.}
\label{fig07}
\end{figure*}

Table~\ref{tab1} compares the number of floating-point operations (FLOPs) and model size of the four networks under different compression ratios. The number of FLOPs and the model size represent the time and space complexity of the model, respectively. As seen from the table, when the compression ratios are $\gamma =$ 32 and 64, PSFNet has the smallest number of FLOPs and model size and has the best NMSE performance. When the compression ratio is $\gamma =$ 32, the number of FLOPs of PSFNet is about 85.2\% and the model size is about 64.1\% of the smallest among the other three methods; when the compression ratio is $\gamma =$ 64, the number of FLOPs of PSFNet is about 24.4\% and the model size is about 31.4\% of the smallest among the other three methods. The number of FLOPs and the size of PSFNet decrease rapidly with the compression ratio increases, which means the PSFNet has both lightweight characteristics and the best NMSE performance compared with the other three methods. In addition, the number of FLOPs and model size of PSFNet decreases rapidly with the compression ratio increase because the compression ratio is related to the number of encoder output channels and decoder input channels. The higher the PSFNet compression ratio, the fewer the corresponding convolution feature channels, thus the corresponding model FLOPs and model size will decrease accordingly. Significant lightweight and impressive reconstruction accuracy make our scheme easier to deploy and apply in actual scenarios than the three benchmark methods.

In addition, since NMSE only characterizes the cumulative element square error between the reconstruction phase shift and the optimal phase shift, it cannot perfectly reflect the detailed recovery effect of a single element of the optimal phase shift. With the matter in mind, we visualize the reconstruction phase shifts of PSFNet and the other three methods under different quantization bits with different compression ratios.

Fig.~\ref{fig07} shows the visualization of the optimal phase shift as the input and the reconstruction phase shift by PSFNet and the other three benchmark methods under different quantization bits when the compression ratios are 16 and 32. The left side of the red dotted line is the optimal phase shift matrix as the input. The right side of the red dotted line is the phase shift matrices reconstructed under different quantization bits. It should be noted that the quantization bit of PSFNet indicates the size of the knowledge base. It can be concluded from Fig.~\ref{fig07} that PSFNet performs a much better effect on the phase shift reconstruction under different compression ratios than the other three methods, especially for restoring reconstruction details at positions with significant phase shift changes. In addition, with the increase of the number of quantization bits, PSFNet has a better reconstruction effect of the feedback phase shift and clearly restores the details of the phase shift matrix. As the number of quantization bits increases, the knowledge base of PSFNet gets larger. Thus corresponding increase of the phase shift information results in higher phase shift reconstruction accuracy. The visual picture with a compression ratio of 64 is referred to the appendix.

\subsection{The Effect of Knowledge Base Learning on Reconstruction Performance}
\begin{figure}[!t]
\centering
\includegraphics[width=3.2in]{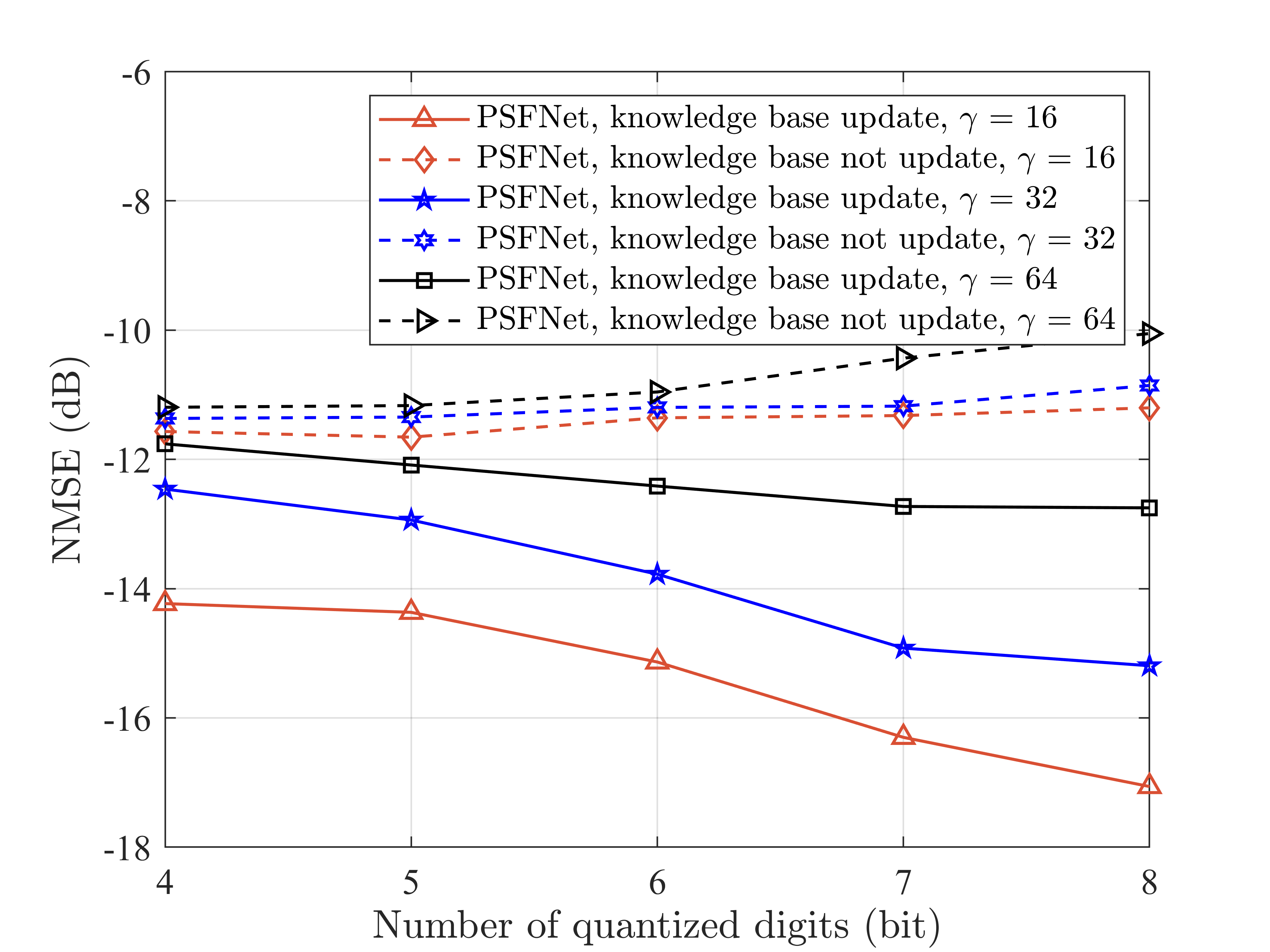}
\caption{Effect of $kb$ loss on PSFNet NMSE performance.}
\label{fig08}
\end{figure}

In Section III $C$, the $kb$ loss enables the knowledge base to learn in the initial training stage so that the distribution of encoder output feature vectors and the knowledge base approximate each other, which is used to represent the update of the knowledge base. In order to confirm the impact of $kb$ loss on network performance, we conducted the following experiments. The solid line in Fig.~\ref{fig08} represents the result using the total loss function, and the dotted line expresses the result of the partial loss function without $kb$ loss. As can be seen from Fig.~\ref{fig08} that under different compression ratios, the NMSE of adding $kb$ loss to the loss function is much smaller than the NMSE of the partial loss function without $kb$ loss, and the model performance will continue improving as the number of quantization bits increasing. It is confirmed that with the addition of $kb$ loss to the total loss function, the knowledge base and the encoder training simultaneously make the distributions of the knowledge base and the encoder output closer. Thus, the output of the encoder is better represented by the limited vectors in the knowledge base, leading to raised phase shift reconstruction accuracy.
% figure 08
% figure 09
\begin{figure}[!t]
\centering
\includegraphics[width=3.2in]{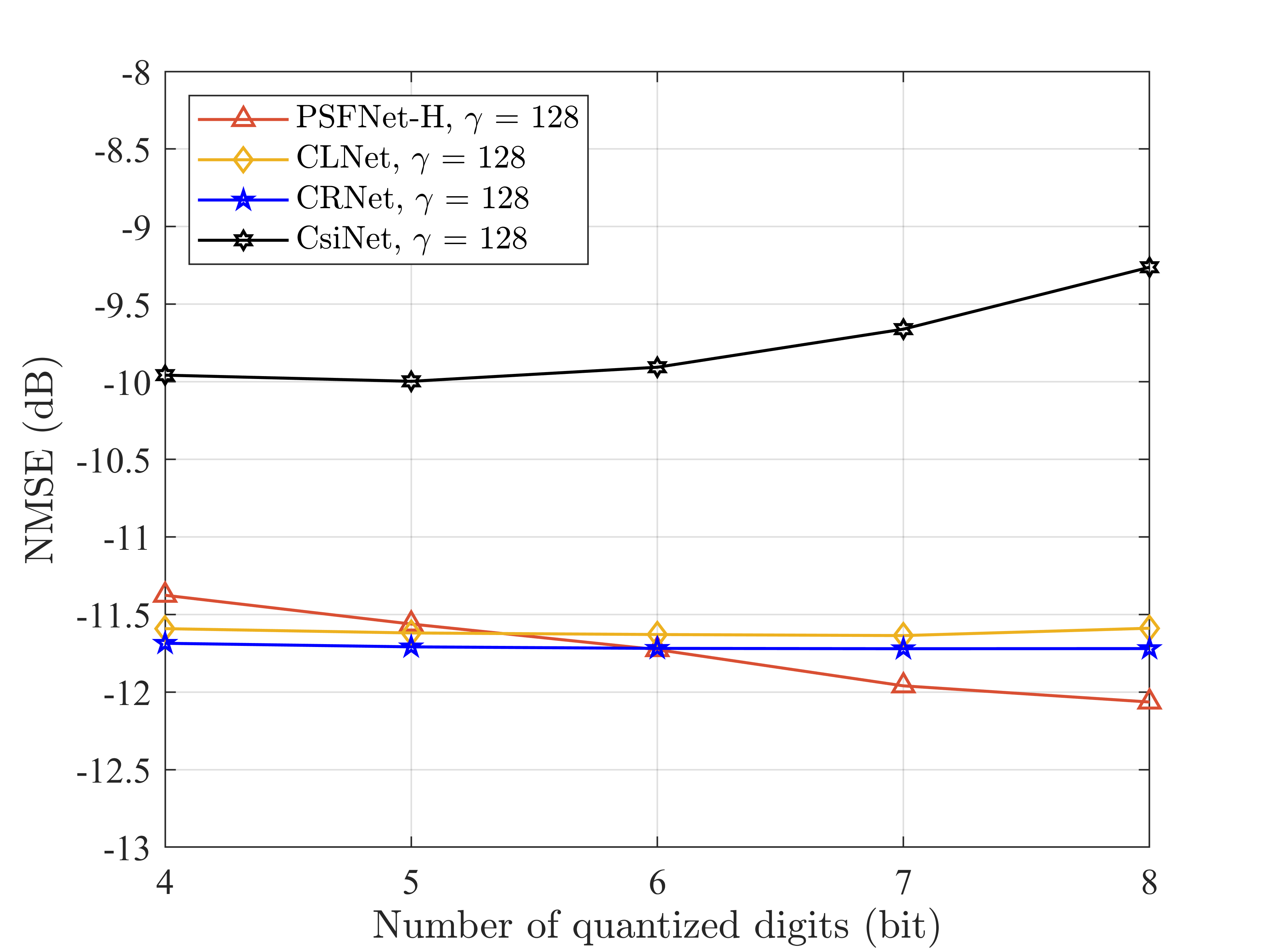}\\
(a)\\
\includegraphics[width=3.2in]{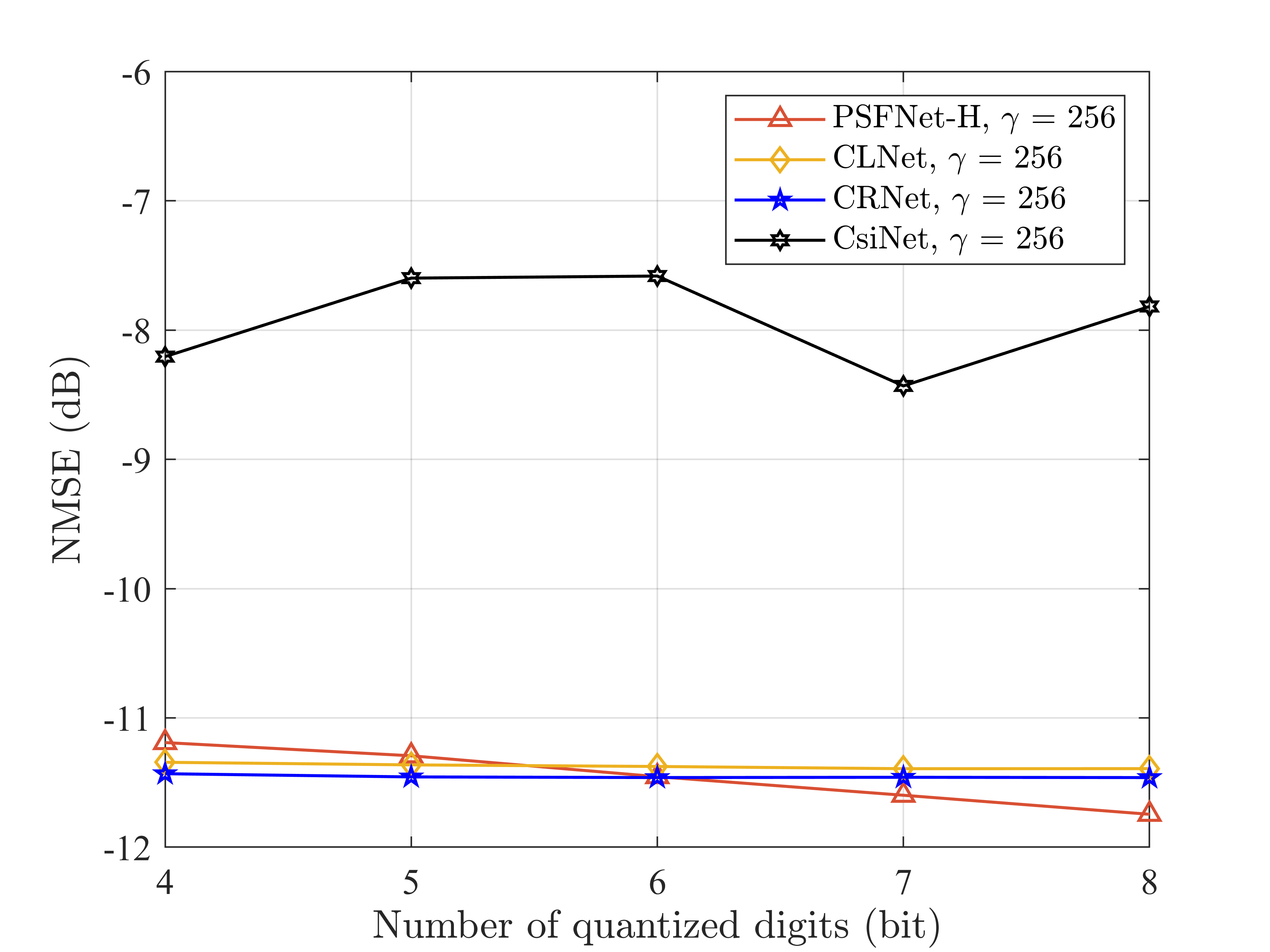}\\
(b)
\caption{NMSE performance comparison of PSFNet-H and other methods under high compression ratios.}
\label{fig09}
\end{figure}

\subsection{PSFNet-H Simulation Results}
For PSFNet-H with a higher compression ratio, the simulation parameters are $N=$ 1024, $M=$ 32, $K=$ 64, $\beta =$ 0.5, $k_0=$ 2, the learning rate is set to 0.003, the batch size is 200, and the number of training epochs is 200. Other simulation conditions are the same as PSFNet.

Fig.~\ref{fig09} shows the NMSE performance comparison among PSFNet-H and CRNet, CLNet, CsiNet under a higher compression ratio. There are different compression ratios $\gamma =$ 128, 256 set in Fig.~\ref{fig09} (a), (b). From Fig.~\ref{fig09}, it can be seen that the NMSE of PSFNet-H decreases with the increase of the number of quantization bits in any case of a higher compression ratio. Moreover, only when the number of quantization bits is 4 or 5, there is a small gap between PSFNet-H and CLNet, CRNet. While as long as the number of quantization bits increases, the NMSE of PSFNet-H is obviously better than the other three methods. This proves that an increase of the knowledge base vectors along with the number of quantization bits improves the accuracy of phase shift reconstruction.
\begin{table*}[!t] 
\caption{Comparison of FLOPs and model size between PSFNet-H and other three benchmark networks.}
\centering
\begin{tabular}{cllll}
\toprule
     Compression ratio   & \multicolumn{2}{c}{$\gamma = 128$} & \multicolumn{2}{c}{$\gamma = 256$}   \\
\midrule
     Model   & FLOPs & Model parameters & FLOPs & Model parameters \\
\midrule
 PSFNet-H (Ours)   & \bf{983.232 K (32.9\%)} & \bf{6.645 K (35.2\%)} & \bf{777.360 K (26.1\%)} 
               & \bf{4.915 K (46.8\%)} \\
CRNet   & 2.992 M & 20.162 K & 2.984 M & 11.966 K  \\
CLNet   & 3.134 M & 20.279 K & 3.126 M & 12.083 K  \\
CsiNet  & 3.049 M & 18.883 K & 3.041 M & 10.687 K  \\
\bottomrule
\end{tabular}
\label{tab2}
\end{table*}

Table~\ref{tab2} compares the number of FLOPs and model size of the four networks under a higher compression ratio. As seen from the table, when the compression ratios are $\gamma =$ 128 and 256, PSFNet-H has the least number of FLOPs and the smallest model size. When the compression ratio is $\gamma =$ 128, the number of FLOPs of PSFNet-H is about 32.9\% and the model size is about 35.2\% of the smallest of the other three methods; when the compression ratio is $\gamma =$ 256, the number of FLOPs of PSFNet-H is about 26.1\% and the model size is about 46.8\% of the smallest among the other three methods. It is worth noting that neither the number of FLOPs nor the model size of PSFNet-H exceeds 1 M under higher compression ratios. Compared with the other three methods, the number of FLOPs and the model size of PSFNet-H are impressively smaller, therefore, PSFNet-H has lightweight features while better NMSE performance.

\begin{figure*}[!t]
\centering
\includegraphics[width=6.2in]{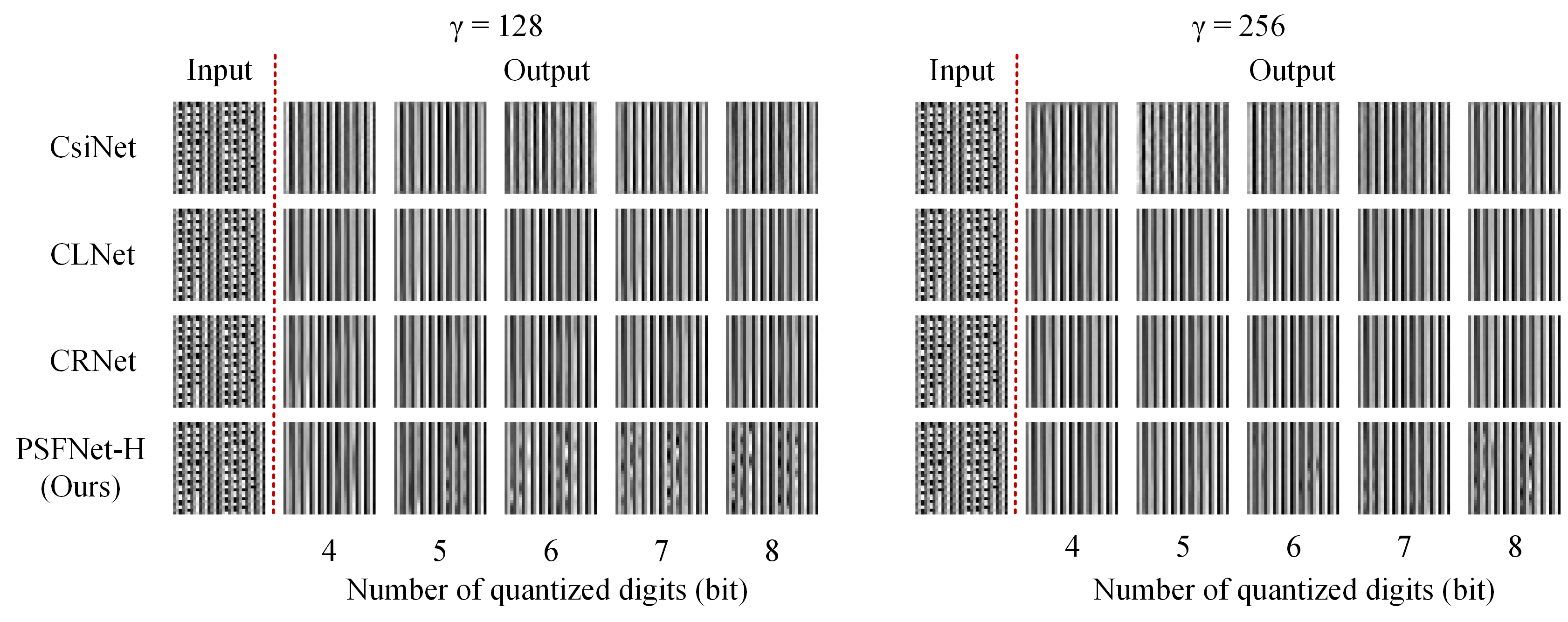}
\caption{Visualization of PSFNet-H and other three benchmark methods reconstructing phase shift.}
\label{fig10}
\end{figure*}

Fig.~\ref{fig10} visualizes the optimal phase shift as the input and the reconstruction phase shift of PSFNet-H and other three benchmark methods with different quantization bits when the compression ratios are 128 and 256. The left side of the red dotted line is the optimal phase shift as the input. The right side of the red dotted line is the phase shift matrices reconstructed under different quantization bits. It can be seen intuitively from the figure that the reconstruction effect of CsiNet is the worst under different quantization digits, and the reconstruction effect of CRNet and CLNet is also unsatisfactory. Due to the high compression ratio, the reconstruction effect of PSFNet-H is predictably worse than PSFNet, nevertheless, it is significantly better than the other three benchmark methods. In conclusion, PSFNet-H has better recovery details and the best phase shift reconstruction effect. In addition, the recovery of phase shift details is promoted gradually with increasing quantization bits.
% figure 10

\section{Conclusions}
In this paper, we first propose a phase shift feedback scheme for mmWave RIS based on the KBAE framework. By establishing a learnable knowledge base at both ends of the feedback, the proposed scheme carries more characteristic information of the phase shift matrix in the limited feedback bit stream and recovers the phase shift matrix more accurately at the RIS. In addition, we propose PSFNet and lightweight PSFNet-H as two autoencoder structures for different compression ratios. Furthermore, we propose the continuous phase shift matrix feedback of RIS for the first time, which improves the efficiency of RIS phase shift feedback. Ultimately, we perform simulations on the proposed RIS phase shift feedback scheme. The simulation results confirm that when the knowledge base is reasonably large, our scheme significantly improves the reconstruction accuracy of RIS phase shift feedback compared with the three benchmark methods, regardless of the compression ratio. Moreover, the proposed scheme performs the minimum time and space complexity and a better visual reconstruction effect.

\section*{Appendix}
When the compression ratio $\gamma $ is 64, PSFNet and other three benchmark methods reconstruct the visualization of the phase shift.
% figure 11
\begin{figure}[H]
\centering
\includegraphics[width=3.2in]{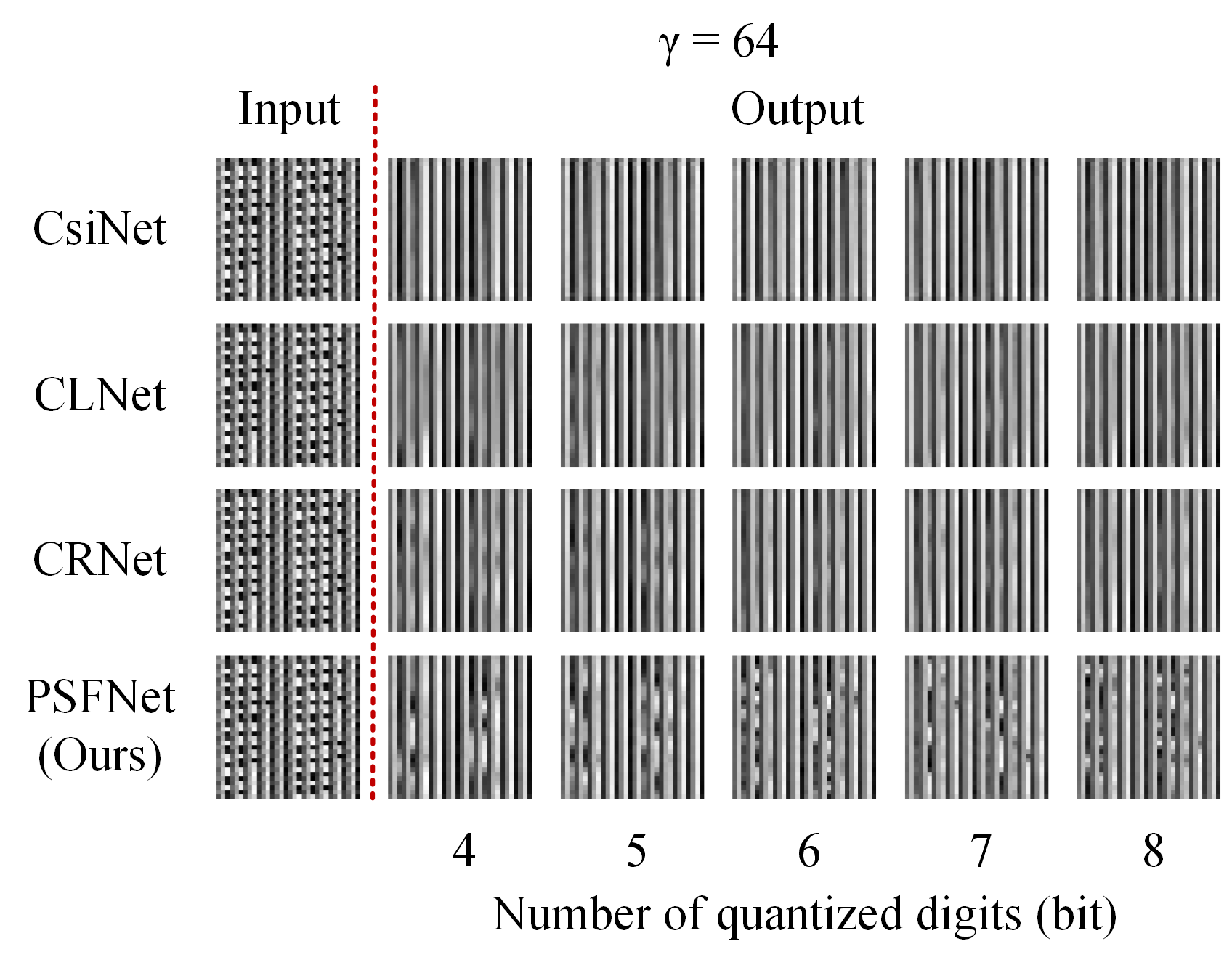}
\caption{Visualization of phase shift reconstruction by PSFNet and other three benchmark methods when $\gamma =$ 64.}
\label{fig11}
\end{figure}

%\section*{Acknowledgments}

\newpage

\vfill

\end{document}